\documentclass[prb,aps,amssymb,twocolumn,showpacs,superscriptaddress]{revtex4-1}
\usepackage{graphicx}
\usepackage{amsmath}
\usepackage{amssymb}
\usepackage[latin1]{inputenc}

\begin{document}

\title{Laser doping for ohmic contacts in n-type Ge}

\author{F. Chiodi}
\email{francesca.chiodi@u-psud.fr}
\affiliation{Institut d'Electronique Fondamentale, CNRS - Universit\'e Paris-Sud, 91405 Orsay cedex, France}
\author{A. D. Chepelianskii}
\affiliation{Laboratoire de Physique des Solides, CNRS - Universit\'e Paris-Sud, 91405 Orsay cedex, France}
\author{C. Gardes}
\altaffiliation{Now at Institut d'Electronique, de Micro\'electronique et de Nanotechnologie, CNRS - Universit\'e Lille 1, 59652 Villeneuve D'Ascq cedex}
\affiliation{Institut d'Electronique Fondamentale, CNRS - Universit\'e Paris-Sud, 91405 Orsay cedex, France}
\author{G. Hallais}
\author{D. Bouchier}
\author{D. D\'ebarre}
\affiliation{Institut d'Electronique Fondamentale, CNRS - Universit\'e Paris-Sud, 91405 Orsay cedex, France}

\date{\today}
\begin{abstract}
We achieved ohmic contacts down to 5 K on standard n-doped Ge samples by creating a strongly doped thin Ge layer between the metallic contacts and the Ge substrate. Thanks to the laser doping technique used, Gas Immersion Laser Doping, we could attain extremely large doping levels above the solubility limit, and thus reduce the metal/doped Ge contact resistance. We tested independently the influence of the doping concentration and doped layer thickness, and showed that the ohmic contact improves when increasing the doping level and is not affected when changing the doped thickness. Furthermore, we characterised the doped Ge/Ge contact, showing that at high doping its contact resistance is the dominant contribution to the total contact resistance.
\end{abstract}
 \maketitle

\emph{Introduction --} Cryogenic detectors are currently used in space imaging to achieve better energy resolution and lower noise levels. However, to further improve the quality of the detection, a close by cryogenic electronics is needed. Silicon JFET, despite their very good performances down to 40 K, are not suitable at lower temperatures due to the carriers freeze-out, so that new materials need to be explored. Germanium JFET are an interesting possibility, having high impedance, low leakage currents and low 1/f noise \cite{Todi,Das}. However, their development is subordinate to the resolution of some technical issues, among which the difficulty in creating ohmic contacts over n-type Ge regions. Indeed, two main problems arise when contacting n-Ge with a metal: on one side, a strong Fermi-level pinning results in a high effective electron Schottky barrier height, independent on the chosen metal; on the other, low solubility and fast dopant diffusion make it difficult to create an intermediate thin layer of strongly doped semiconductor, as is often done to achieve ohmic contacts. 
Several approaches have been validated to realise ohmic, low resistance contacts on n-Ge, such as the suppression of the Fermi level pinning by the introduction of ultrathin SiN \cite{Kobayashi} and Ge$_3$N$_4$ \cite{Lieten} layers or by sulfur passivation \cite{Thathachary}. Alternatively, Sb $\delta$-doping \cite{Sawano}, dopant segregation during Ni germanidation \cite{Firrincieli2} and laser annealing \cite{Wang,Firrincieli} were successfully used to obtain high doping concentrations and ohmic contacts. Because of the higher dopant activation, RTA (rapid thermal annealing) \cite{Chui,Thareja} and more recently laser annealing \cite{Wang} produced the higher concentrations and the lower contact resistances. Indeed, as the laser pulse duration is extremely short ($\sim 20\,$ns), the dopants don't have the time to diffuse to their equilibrium concentration, and it is thus possible to attain a meta-stable state where the active concentration is higher than the solubility limit. Between the laser annealing techniques, Gas Immersion Laser Doping (GILD) has a particular interest, as it doesn't require an implantation step prior the laser annealing. The samples are thus free of all deep defects introduced by the ion implantation, which cannot be completely erased even by fast laser annealing. Moreover, GILD gives the possibility to dope a specific area of a device and, without breaking the vacuum, to add a thin ultra-doped layer on top to insure the ohmic contact, eliminating the need for further processes.\\
In this paper, we demonstrate ohmic contacts on standard n-doped Ge samples of concentration $10^{15}\,$ cm$^{-3}$, by creating between the metallic contact (M) and the Ge substrate (nGe) a strongly Phosphorus doped thin Ge layer (n$^{++}$Ge) of resistivity $\rho_{++}$ down to $5\times 10^{-4} \,\Omega \, cm$. We tested the influence of the doped layer thickness $t_{++}$ and doping concentration $n_{++}$ on the M/n$^{++}$Ge contact resistance $\rho_{c,M}$. We then characterised the nGe/n$^{++}$Ge contact, showing that its contact resistance $\rho_{c,n}$ is independent on doping thickness and concentration. The distinction between the different contributions of the contact resistance has been often overlooked in the literature, but shouldn't be neglected: indeed, we showed that $\rho_{c,n}$ dominates the total contact resistance as soon as $\rho_{c,M}$ is better than $2 \times 10^{-3} \, \Omega \,$cm$^2$. \\
In view of the utilisation of Ge-based devices such as Ge JFET at low temperature, we tested the I-V curves down to 5 K, and confirmed that the I-V characteristics remain ohmic even at low temperatures. The total contact resistance $\rho_{c,t}$ was also extracted, showing that the contact is better than at room temperature down to 13 K, with a minimum resistance at 40 K where $\rho_{c,t} (T)/\rho_{c,t}(300K)=13 \,\%$.

\begin{figure}
\includegraphics[width=0.8\columnwidth]{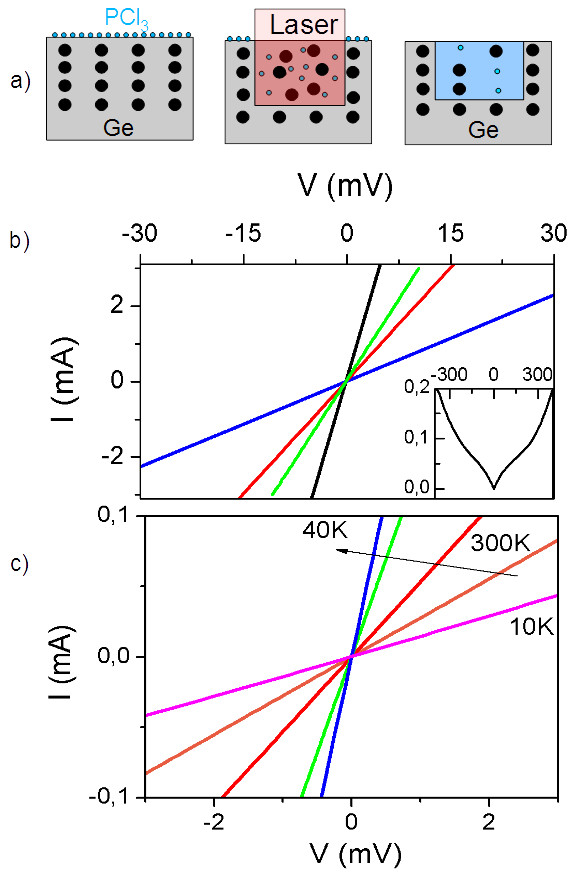}
\caption{(color online)
a) GILD process: chemisorption of the precurson gas over the sample surface; melting of the substrate by the laser pulse and P introduction in the liquid Ge; fast cooling and epitaxy of a Ge:P layer on top of the Ge substrate. b) Representative $I(V)$ curves for as-made samples with different doping concentrations (blue and green: 10 and 50 laser shots, $t_{++}=$85 nm; red and black: 5 and 10 laser shots, $t_{++}=$175 nm) 
%black: $1.1\times 10^{20}\,$cm$^{-3}$, 85 nm; green: $2.7\times 10^{19}\,$cm$^{-3}$, 175 nm; red: $5.5\times 10^{20}\,$cm$^{-3}$, 85 nm; blue: $5.4\times 10^{20}\,$cm$^{-3}$, 175 nm). 
Inset: Non linear $|I|(V)$ curves for $t_{++}=$175 nm and 1 laser shot, after the $n_{++}$ layer etch between the contacts; units are the same as in the main frame.
%(5.4$\times 10^{18}$cm$^{-3}$)
 c) Ohmic $I(V)$ characteristics at 300K (orange), 200 K (red), 100 K (green), 40 K (blue), 10 K (pink) for an etched sample with $t_{++}=$175 nm and 5 laser shots ($n_{++}=2.7 \times 10^{19}$cm$^{-3}$). The voltage of the curve at 10 K has been rescaled by a factor of 10. }
\label{VI}
\end{figure}

\emph{Laser doping and sample fabrication--} The quality of the realised interfaces rests on the many advantages of the adopted doping technique, Gas Immersion Laser Doping (GILD). The doping takes place in a ultra high vacuum chamber, at a base pressure of $10^{-9}\,$mbar which insures a very low impurity level. A puff of the precursor gas, PCl$_3$, is injected using a pulse valve onto the Ge sample surface, where it saturates the chemisorption sites. The gas is continuously pumped and, after a small delay, the substrate is melted by a pulsed excimer XeCl laser, of pulse duration 25 ns. The phosphorous can thus diffuse into the liquid germanium phase and is incorporated in the lattice as the liquid/solid interface moves back to the surface at the end of the irradiation. A Ge:P crystal is thus created by liquid phase epitaxy above the underlying Ge substrate (Fig. \ref{VI}-a). \\
Because of the short pulse duration and the high recrystallization speed, high phosphorous concentrations, larger than the solubility limit ($\sim$ 1$\times 10^{19}$cm$^{-3}$) \cite{Harame}, can be obtained by multiple process repetitions \cite{Bhaduri,Cammilleri}. As the dose of active dopants is determined exclusively by the number of laser shots, the doping depth can be independently tuned by controlling the laser energy. Above the Ge melting threshold ($\sim 250\,$mJ/cm$^2$), we can thus linearly increase the melted depth up to a few hundreds of nanometers. To monitor the doped depth, we measure the in-situ time resolved transient reflectivity at 675 nm for each laser shot. Since the reflectivity changes in liquid Ge, we can directly measure the melting duration, which is proportional to the doping depth \cite{Kerrien2003}.
To improve the uniformity of the $2\,$mm$\times2\,$mm laser spots, the spatial inhomogeneity of the laser energy density was reduced to less than $1\%$ by a careful optical treatment of the laser beam (using, in particular, a fly eye homogenizer). This results in a constant P concentration in the doped volume and a sharp 2D Ge:P/Ge interface, as demonstrated by SIMS measurements \cite{Cammilleri}. 

All the laser-doped samples were fabricated on a single n-type 101 Ge wafer of doping concentration $10^{15}\,$ cm$^{-3}$. Two different thicknesses were studied, $t_{++}$=85 and 175 nm (laser energy density of 450 and 640$\,$mJ/cm$^2$ respectively), and for each a varying number of laser shots (1-5-10-50) determined the doping level. This resulted in the concentration range $n_{++}=5\,\times 10^{18}- 1\,\times 10^{20}\,$cm$^{-3}$. Three identical set of samples were realised for each condition. 
After a surface desoxidization, Ti (15 nm)/Al (200 nm) rectangular contacts were deposited in a Transmission Line Method (TLM) configuration by e-beam evaporation. Ohmic M/nGe contacts were also demonstrated for Ni/Au and Ti/Au contacts, stressing the robustness of the obtained results. 
   
\emph{Results --} Resistance measurements were performed on the M/n$^{++}$Ge/nGe samples with a set of 9 rectangular TLM. Ohmic $I-V$ characteristics were consistently obtained (Fig. \ref{VI}-b), proof of the good quality of the contact realised. The resistance, dominated by the doped layer resistance, decreased when increasing the doping level $n_{++}$ or the doped thickness $t_{++}$. Only in the particular case of one laser shot, rectifying I-V characteristics were found for both doping thicknesses (Fig. \ref{VI}-b, inset). This could be attributed to the imperfect dopant diffusion in the melted layer induced by a single laser shot, as was already shown in Si:B laser doped systems \cite{Bhaduri}, confirming the necessity of an efficient doping process. 
\begin{figure}[t!]
\includegraphics[width=\columnwidth]{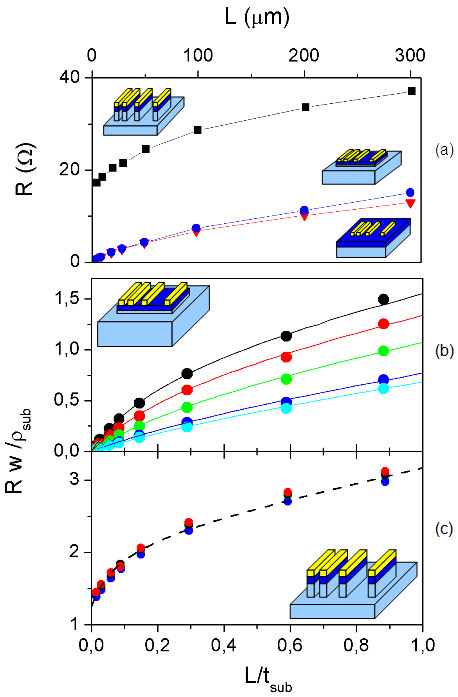}
\caption{(color online) a) $R(L)$ curves for $n_{++}=$1.1$\times 10^{20}$cm$^{-3}$ and $t_{++}=$85 nm as realised (red triangles), after the trench etch around the contacts (blue dots) and after the deep etch through the doped layer (black square). b) Simulations and dimensionless $w R(L)/\rho_{sub}$ data for pristine samples after the trench etch and c) after the deep etch through the doped layer. Symbols from top to bottom on pannel b): black: $n_{++}=5.6\times 10^{19}\,$cm$^{-3}$, $t_{++}$=85 nm; red: $n_{++}=1.1\times 10^{20}\,$cm$^{-3}$, $t_{++}$=85 nm; green: $n_{++}=2.7\times 10^{19}\,$cm$^{-3}$, $t_{++}$=175 nm; blue: $n_{++}=5.4\times 10^{19}\,$cm$^{-3}$, $t_{++}$=175 nm; light blue: $n_{++}=2.7\times 10^{20}\,$cm$^{-3}$, $t_{++}$=175 nm. Lines represent our best-fitting numerical simulations. The same notations apply to c).
}
\label{RL}
\end{figure}

The contact resistance consists of two independent contributions given by the M/n$^{++}$Ge and n$^{++}$Ge/nGe contact resistances $\rho_{c,M}$ and $\rho_{c,n}$. To gain access to $\rho_{c,n}$, we removed the doped Ge layer between the contacts by anisotropic reactive ion etching.  The Al contacts were unaffected by the CHF$_3$ based etch and acted as an etching mask, preserving the doped layer intact under the contacts. In the exposed areas between the contacts the etch was deep enough to completely remove the n$^{++}$Ge, penetrating up to $1.6\,\mu$m into the nGe substrate. The resulting samples also had linear $I-V$ characteristics, proving that ohmic contacts are obtained both at the M/n$^{++}$Ge and at the n$^{++}$Ge/nGe interfaces. 
We measured the temperature dependence of the $I-V$ characteristics of the etched samples (Fig. \ref{VI}-c) and found that, independently of the doping concentration, the contacts remain ohmic down to 5 K. 
To separate the contribution of the metal/doped Ge interface from the doped Ge/Ge interface, we compared the resistance $R$ of pristine and etched samples for different separations $L$ between the TLM contacts. The room temperature dependence $R(L)$ measured for the two types of samples is shown on Fig. \ref{RL}-a. 
We attribute the non-linearity of the $R(L)$ dependence to the finite ratio between the substrate thickness $t_{sub}$=$340\, \mu$m and the TLM distance $L=5-300 \,\mu$m. Indeed the $R(L)$ dependence of the etched samples shows a more pronounced non-linearity, which can be easily understood by considering that in the unetched bilayers most of the current is transmitted through the thin doped layer, satisfying in part the thin film approximation leading to a linear $R(L)$.
To confirm that current paths at the end of the contacts do not contribute significantly to the $R(L)$ dependence, we prepared a deep trench surrounding the TLM contacts. As shown on Fig.~\ref{RL}-a this procedure left the $R(L)$ dependence unchanged except at the highest $L$ values. This observation is consistent with the small ratio between $L$ and the width of the contacts $w = 800\;{\rm \mu m}$ in our samples. In our future discussion, we will indicate as pristine samples the samples surrounded by a trench, with the doped film intact between the contacts.
We performed a systematic investigation of the $R(L)$ dependence for pristine and etched samples prepared with different doping concentrations of the n$^{++}$Ge layer  (see Figs. \ref{RL}-b,c). For the etched samples the doping level can only change the offset at $L=0$ in the $R(L)$ dependence. Our measurements demonstrate that the offset is actually almost independent of the doping concentration. Also, the offset increases significantly after etching for all the samples. These two observations suggest that the n$^{++}$Ge/nGe contact resistance is the dominant contribution to the total contact resistance and that it is almost independent of the doping in the n$^{++}$Ge layer. To further confirm this qualitative analysis and to extract quantitative parameters from our measurements we simulated the shape of the non-linear $R(L)$ curves by solving numerically the Laplace equations on the electrostatic potential inside the bilayer using a finite elements method. The simulations were performed in a two dimensional geometry where the effects of the finite contact width ($w = 800\;{\rm \mu}$m) were neglected. This model allowed us to take into account the geometrical effects due to the finite thickness of the Ge sample, and to describe the current exchange between the n$^{++}$Ge and nGe layers in presence of a finite contact resistance between the layers. 

As demonstrated by the quality of the fits in Fig. \ref{RL}-b and \ref{RL}-c we managed to obtain a very good quantitative description of the measured $R(L)$ dependences. The $R(L)$ curves of the etched samples were fitted with a single adjustable parameter corresponding to a substrate resistivity $\rho_{sub} = 0.95 \,\Omega cm$, in good agreement with the manufacturer value $\rho_{sub} = 1-1.3 \,\Omega cm$. The total contact resistance $\rho_{c,t}$ could then be deduced from the resistance offset at $L=0$ in the $R(L)$ dependences. 
The analysis of the $R(L)$ curves for the pristine samples involved two fitting parameters, the doped layer square resistance $R_{sq} = \rho_{++}/t_{++}$ and $\rho_{c,n}$. The contact resistance $\rho_{c,M}$ was deduced as previously from the offset at $L=0$. 
We checked the validity of our fitting procedure by measuring in a Van der Pauw configuration the doped layer square resistance $R_{sq}$ of a second set of identical laser spots with no metallic contacts, finding a relative error with the fitting results within $\sim$10$\%$.
Moreover, the sum of $\rho_{c,M}$ and $\rho_{c,n}$ extracted from the resistance of the pristine samples reproduced the values of the total contact resistance $\rho_{c,t}$ measured on the etched samples to within $5\%$. We are thus confident that the combination of our experimental results and numerical simulations allows us to measure the contact resistances at both the n$^{++}$Ge/nGe and M/n$^{++}$Ge interfaces.
 
\begin{figure}
\includegraphics[width=0.9\columnwidth]{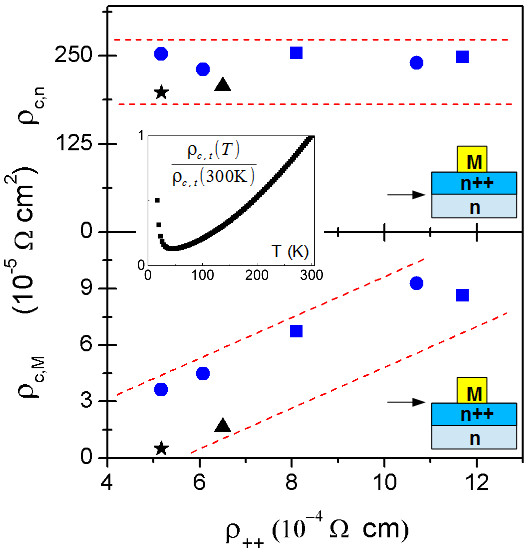}[t]
\caption{(color online) Contact resistances $\rho_{c,n}$ and $\rho_{c,M}$ as a function of the doped layer resistivity $\rho_{++}$. Top panel: $\rho_{c,n}$, broadly independent of the resistivity. Bottom panel:  $\rho_{c,M}$, decreasing linearly with the decrease of the doped layer resistivity (increasing concentration) down to a minimum of $5 \times 10^{-6} \, \Omega cm^2$. Two different series of samples are shown: the blue symbols (squares for $t_{++}$=85 nm and dots for $t_{++}$=175 nm) and the black symbols (triangle for $t_{++}$=85 nm and star for $t_{++}$=175 nm); originally identical, the latter had a more through desoxidation. Lines are guides to the eye. Inset: total contact resistance $\rho_{c,t}$ normalised to its value at 300 K, as a function of temperature. }
\label{Rc}
\end{figure}

In Fig. \ref{Rc} we plot the extracted contact resistances $\rho_{c,M}$ and $\rho_{c,n}$ against the resistivity of the laser doped layer
$\rho_{++}$. Two sets of samples are presented; a more thorough deoxydation of one set allowed us to reduce $\rho_{c,M}$ down to around  $5\,\times 10^{-6} \,\Omega \,$cm$^2$, a value above but comparable with state of the art results ($\sim 10^{-6} - 5\times 10^{-7}\, \Omega \,$cm$^2$) \cite{Firrincieli2,Gallacher}. We  observe that $\rho_{c,M}$ decreases linearly with $\rho_{++}$ producing a better ohmic M/n++ interface at higher doping concentration (corresponding to lower $\rho_{++}$ values). On the other hand, the contact resistance at the n$^{++}$Ge/nGe interface is found to be independent on the doping level in the n++ layer. Thus at high dopings, $\rho_{c,n}$ gives the dominant contribution to the total contact resistance. This highlights the importance of the n$^{++}$Ge/nGe interface whose contribution has so far been neglected, as to our knowledge previous experiments measured only the contact resistance at the M/n$^{++}$Ge interface.
In view of the cryogenic electronics applications of our method, we
measured the temperature dependence of the resistance between two
different TLMs ($L$=200 $\mu$m and $L$=300 $\mu$m). The characteristic extracted total contact resistance $\rho_{c,t}$ decreases monotonically with the temperature down to 40 K, where its value is only 13 $\%$ of its room temperature value (Fig. \ref{Rc}, inset). A sharp increase is measured at lower temperatures due to the diminished carrier density activation; however, at 16 K, the contact is still only half of its value at room temperature. A behaviour similar to the one we observe above 50 K was reported for the contact resistance decrease of strongly doped Si:P/metal layers \cite{Swirhun}. 

\emph{Conclusions --} In conclusion, we have shown that ohmic contacts down to 5 K can be consistently realised on a n-doped Ge substrate by strong laser doping of a thin surface layer. We have measured the contact resistance between the Ge substrate and the metallic electrodes, separating the contributions of the M/n$^{++}$Ge and of the n$^{++}$Ge/nGe interfaces. While $\rho_{c,n}$ was shown to be independent on the thickness and concentration of the laser doped layer, we could strongly suppress $\rho_{c,M}$ by increasing the doping level.  
The GILD laser doping technique was fundamental in creating the high doping of the thin epitaxied layers ensuring a good contact resistance. Since GILD also allows to realise the doping in the active-device regions this technique may be crucial for cryogenic temperatures Ge-based electronics.

We are grateful to J.-L. Perrossier  for technical help and discussions.
This work was supported in part by the R\'eseau RENATECH and European FP7 grant No. 263455.

\end{document}